\documentclass[twocolumn,showpacs,secnumarabic,amssymb, aps, prb, reprint]{revtex4-1}
\usepackage{graphicx}
\usepackage{dcolumn}
\usepackage{epsfig}
\usepackage{amsmath}

\begin{document}

\newcommand{\etal}[0]{\textit{et al.}}
\newcommand{\alumina}[0]{Al$_2$O$_3$}
\newcommand{\Kalumina}[0]{$\kappa$-Al$_2$O$_3$}
\newcommand{\Aalumina}[0]{$\alpha$-Al$_2$O$_3$}
\newcommand{\Galumina}[0]{$\gamma$-Al$_2$O$_3$}
\newcommand{\bea}[0]{\begin{eqnarray}}
\newcommand{\eea}[0]{\end{eqnarray}}
\newcommand{\nn}[0]{\nonumber}
\newcommand{\mtext}[1]{\mbox{\tiny{#1}}}
\renewcommand{\vec}[1]{{\bf #1}}
\newcommand{\op}[1]{{\bf\hat{#1}}}
\newcommand{\galpha}[0]{$\alpha$}
\newcommand{\gbeta}[0]{$\beta$}
\newcommand{\ggamma}[0]{$\gamma$}
\newcommand{\gdelta}[0]{$\delta$}
\newcommand{\gepsilon}[0]{$\eps ilon$}
\newcommand{\gphi}[0]{$\phi$}
\newcommand{\gvarphi}[0]{$\varphi$}
\newcommand{\geta}[0]{$\eta$}
\newcommand{\gtheta}[0]{$\theta$}
\newcommand{\gomega}[0]{$\omega$}
\newcommand{\gchi}[0]{$\chi$}
\newcommand{\gxi}[0]{$\xi$}
\newcommand{\gkappa}[0]{$\kappa$}
\newcommand{\del}[1]{\partial_{#1}}
\newcommand{\av}[1]{\langle #1\rangle}
\newcommand{\bra}[1]{\langle #1\right|}
\newcommand{\ket}[1]{\left| #1\right\rangle}
\newcommand{\bracket}[3]{\langle #1 | #2 | #3\rangle}
\newcommand{\sproduct}[2]{\langle #1 | #2\rangle}
\newcommand{\Bracket}[3]{\left\langle #1 \left| #2 \right| #3\right\rangle}
\newcommand{\sign}[1]{\mbox{sign}\left(#1\right)}
\newcommand{\mc}[3]{\multicolumn{#1}{#2}{#3}}
\newcommand\T{\rule{0pt}{2.6ex}}
\newcommand\B{\rule[-1.2ex]{0pt}{0pt}}

\title{\textit{Ab-initio} thermodynamics of deposition growth:
surface terminations of CVD titanium carbide and nitride
}

\author{Jochen Rohrer}
\email{rohrer@chalmers.se}
\affiliation{
BioNano Systems Laboratory, 
Department of Microtechnology, 
MC2, 
Chalmers University of Technology, 
SE-412 96 Gothenburg
}
\author{Per Hyldgaard}
\affiliation{
BioNano Systems Laboratory, 
Department of Microtechnology, 
MC2, 
Chalmers University of Technology,
SE-412 96 Gothenburg
}

\date{\today}

\begin{abstract}
We present a calculational method to predict 
terminations of growing or as-deposited surfaces
as a function of the deposition conditions.
Such characterizations are valueable for understanding catalysis
and growth phenomena.
The method combines \textit{ab-initio} density functional theory (DFT) calculations
and experimental thermodynamical data
with a rate-equations description of partial pressures in the
reaction chamber.
The use of rate equations enables a complete description
of a complex gas environment in terms of a few,
(experimentally accessible) parameters.
The predictions are based on comparisons between free energies of reaction 
associated with the formation of surfaces with different terminations.
The method has an intrinsic non-equilibrium character.
In the limit of dynamic equilibrium (with equal chemical potential in  
the surface and the gas phase)
we find that the predictions of the method
coincide with those of standard \textit{ab-initio} based
equilibrium thermodynamics.
We illustrate the method for chemical vapor deposition (CVD) of TiC(111) and TiN(111), 
and find that the emerging termination can be controlled both
by the environment and the growth rate.
\end{abstract}

\maketitle

\section{Introduction}
The structure and chemical composition of 
surfaces play a fundamental role
in many industrial applications.
Identifying and ultimately learning to control
surface terminations is a key issue
in modern materials design.
In  heterogenous catalysis \cite{ReactionRates}
chemical reactions typically take place
on the solid surface of the catalyst.
Surfaces with different chemical compositions,
for example, different surface terminations,
support different chemical reactions.\cite{Weiss20021,CamilloneIII2002267,Schoiswohl}
This is true
even if the exposed surfaces possess
identical crystallographic orientations.
Similarly, surface terminations strongly influence 
nucleation and growth processes.
Different surface terminations do, in general,
favor adsorption of different atomic species\cite{Kitchin200566,Lodziana:11261,VojvodicLiC}
and therefore affect subsequent  
growth of multicomponent materials.

Atomistic modeling\cite{TLEinstein,AM_Scheffler,AM_Norskov,AM_MYT,AM_OTHERS}
is highly valuable for design of functional materials and surfaces.
It serves as an important complement to and extension of experimental characterizations.
It also provides predictive power and thus an opportunity to accelerate
innovation.\cite{innovation}
Coupling DFT to either thermodynamic\cite{PhysRevB.35.9625,PhysRevB.62.4698,AIT_Scheffler,PhysRevB.70.024103,Tan199649}
or kinetic modeling\cite{kMC_Kratzer,kMC_Bogicevic,PhysRevB.72.115401,PhysRevLett.83.2608}
has proven extremely useful for interpreting and complementing experimental techniques
in characterization of surfaces,
\cite{surfaces}
thin films,\cite{thinfilms}
and interfaces.\cite{interfaces}
Such modeling is often well-suited for descriptions of structures
that are fabricated or investigated 
under well-controlled conditions,
such as ultra-high vacuum (UHV) or molecular beam epitaxy\cite{Arthur2002189} (MBE).

Industrial mass production of materials often employ more efficient
but less-controlled deposition methods,
for example, chemical vapor deposition (CVD). \cite{CVD}
Such methods give rise to compositions and structures
that need not be stable
in a thermodynamical sense under ambient or 
ultra-high vacuum (UHV) conditions.
The fast deposition causes a lot more complexity in the growth proccess
than is found in MBE.
For example, it causes 
cluster formation and formation of intermediates 
in the gas phase or on the substrate
and these structures define barriers for subsequent deposition events.
Relevant experimental characterizations of surface terminations
in such cases are difficult 
and require in-situ measurements.
Atomistic, \textit{ab initio} based modeling
is consequently of even higher value for as-deposited structures.
However, a simple scale-up of the
kinetic modeling\cite{kMC_Kratzer,kMC_Bogicevic,PhysRevB.72.115401,PhysRevLett.83.2608}
faces enormous practical problems
because complex, unknown, and constantly evolving transition states
determine the kinetics of the deposition.

In this paper, we propose an \textit{ab-initio} thermodynamics method
for deposition growth (hereafter referred to as AIT-DG).
Our method  combines DFT calculations with thermodynamic
concepts and rate equations.
The latter effectively describe 
supply and exhaust of gases to and from a
reaction chamber as well as 
deposition processes inside.
Steady-state solutions of these rate equations
enable us to determine Gibbs free energies of reaction
\cite{Demirel} 
of deposition processes 
that lead to different surface terminations.
The free energies of reaction
allow us to predict surface
terminations as a  function of the 
growth environment.

We emphasize that the AIT-DG method is fundamentally different from 
the \textit{ab-initio}  thermodynamics method
proposed in Refs.~\onlinecite{PhysRevB.62.4698}
and \onlinecite{AIT_Scheffler},
also adapted to interfaces,\cite{PhysRevB.70.024103}
and used by many others.
Such methods are designed to predict terminations 
of surfaces that are in \textit{equilibrium} with a given gas environment 
(and are hereafter referred to as AIT-SE, 
\textit{ab-initio} thermodynamics with surface equilibrium).
In contrast, our AIT-DG formulation possesses a  non-equilibrium character,
but formally contains the AIT-SE method 
in the regime, sometimes called dynamic equilibrium,\cite{Tschoegl}
where one can assume equality of the 
chemical potentials in the surface and in the gas.

We illustrate the AIT-DG method for the CVD 
TiX(111) (X = C or N).
These belong to the simplest class of materials 
where different surface terminations can arise.
At the same time there is a high industrial
interest in these materials.
TiN is widely used as diffusion barrier
for Al-based interconnects \cite{li:583}
and finds application as ohmic contact in GaN semiconductor
technology. \cite{0953-8984-12-49-331}
TiC can be used  as substrate for growth 
of other carbidic materials such as SiC
or graphene,\cite{Hillel1993183}
and its (111) surface also shows a potential as catalyst. \cite{vojvodic:146103}
Moreover, TiC and TiN both are commonly used  
as TiX/alumina  multilayer coatings in cutting tool industry. \cite{Halvarsson1993177}

The paper is organized as follows.
Section~\ref{sec1} focuses on the discussion of
surface terminations of binary materials.
In Section~\ref{sec:AIT-DG},
we present our AIT-DG method in detail,
exemplified for CVD of TiX,
and also considering the equilibrium limit of our method.
Section~\ref{sec4} summarizes the computational
method that we use for all \textit{ab-initio} calculations.
We present our results in Section~\ref{sec5}
and discuss these as well as the method itself
in Section~\ref{sec6}.
Section~\ref{sec7} contains a summary
and our conclusions.

\section{Surface terminations\label{sec1}}
\begin{figure}
\begin{tabular}{c}
\epsfig{file=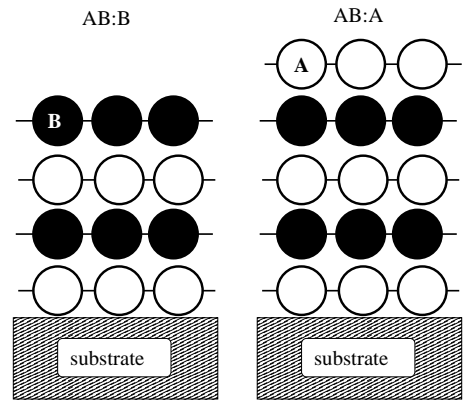,width=6.0cm}
\end{tabular}
\caption{
\label{fig:SurfaceComposition}
Different surface terminations for a film deposited 
on a substrate.
The schematics shows the case for a binary material AB
formed of alternating A  and B layers,
represented by differently colored balls.
While the interfacial composition
largely is determined by the substrate properties,
the  composition of the surface layer
strongly depends on the deposition environment,
possibly leading to an AB film that slightly
deviates from full stoichiometry.
}
\end{figure}

Figure~\ref{fig:SurfaceComposition} shows a schematics of 
thin films of a binary material AB 
(deposited on a substrate),
with the panels  AB:A and AB:B
differing essentially only by their surface terminations.
Here and in the following, 
AB:A (AB:B) denotes an AB surface with A-termination (B-termination).
The figure illustrates the character of our modeling
where the difference in surface terminations
is represented by films of different thicknesses;
formally, the left panel represents a stoichiometric film
grown to $2n$ layers,
while the right panel shows a nonstoichiometric film
grown to $2n+1$ layers.\cite{stoichiometry}
This case represents the simplest case of
surfaces that can have more than one termination,
and the question arises which one will be
created in a specific fabrication process.

Well-defined surface terminations can be created in a variety
of different deposition methods,
including the well-controlled MBE
and less-controlled methods,
such as CVD.
MBE employs UHV
and allows, in principle,
one atom (or dimer) to be deposited at a time.
An abrupt termination of the growth process
therefore provides good explicit control over
surface coverage and termination.

Less-controlled deposition methods typically
yield much higher growth rates and are therefore 
more popular for large-scale manufacturing.
In CVD, a substrate is exposed 
to an inflowing  supply-gas mixture  at high temperatures
and at, compared to MBE, high pressures.
Since there is no explicit control on the number of individual
atoms or molecules in the  gas environment or on the surface,
one cannot simply end the deposition process 
so as to create a pre-specified surface termination.
A more involved analysis is necessary to 
understand what termination will emerge 
in deposition growth.

\begin{figure}
\begin{tabular}{c}
\epsfig{file=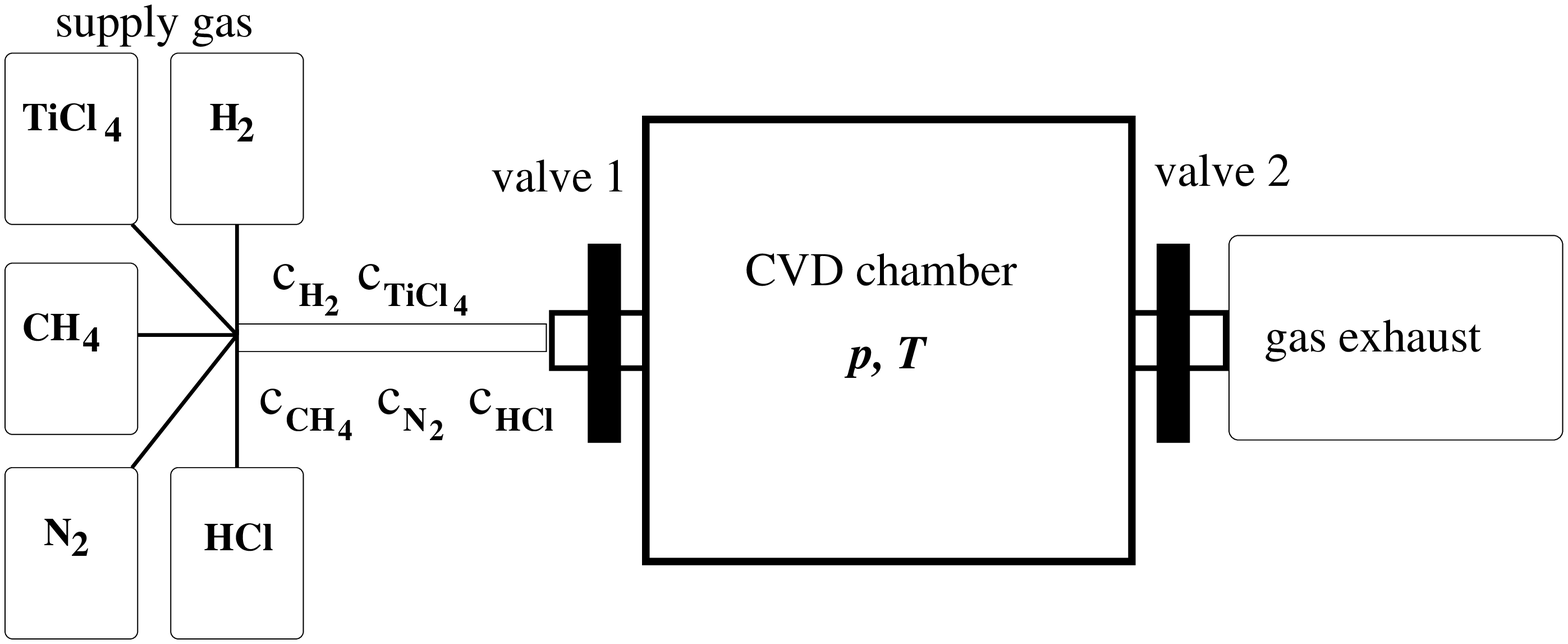,width=8.0cm}\\[0.3cm]
\epsfig{file=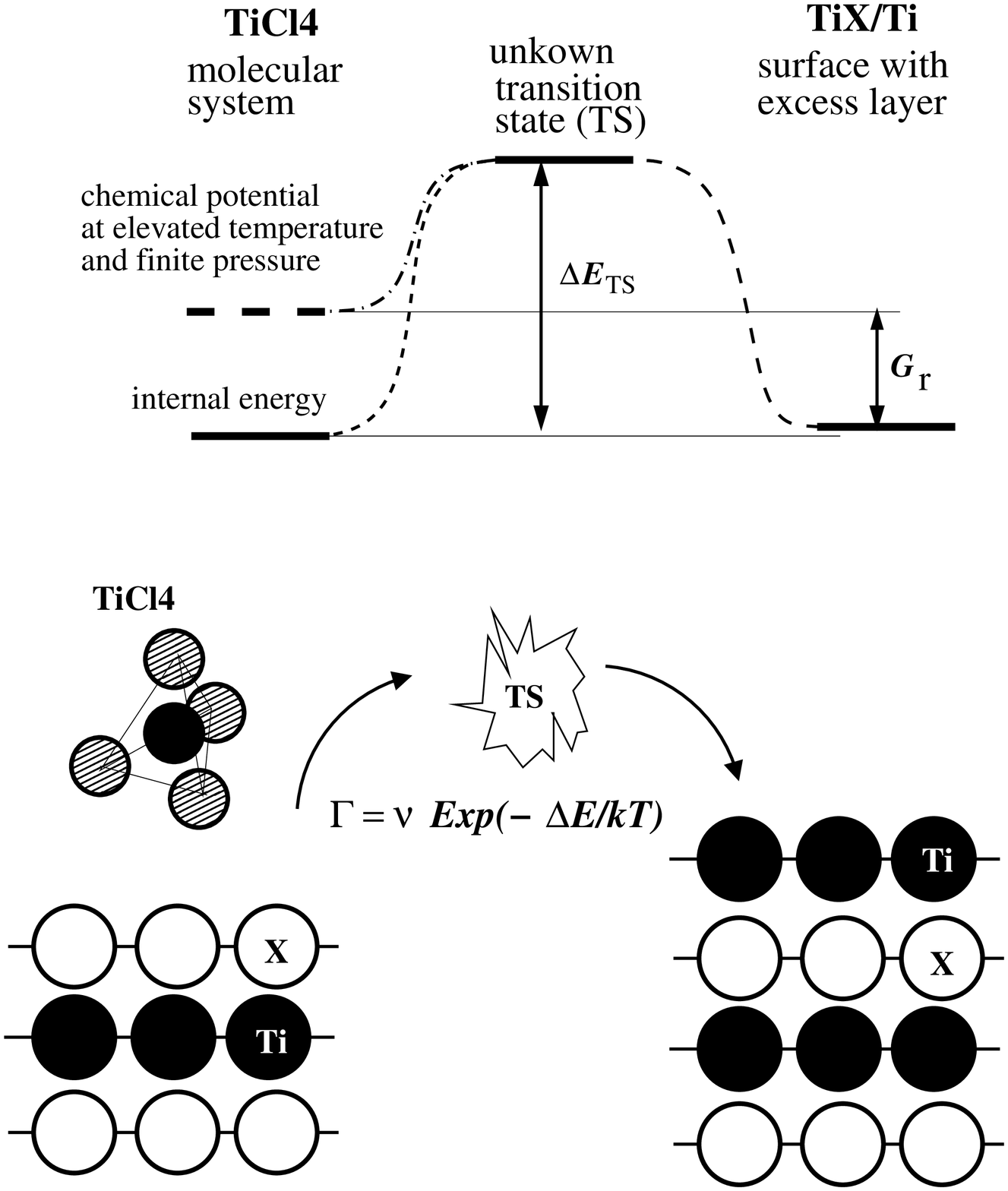,width=6.8cm}\\
\end{tabular}
\caption{
\label{fig:CVDschematics}
Schematics of chemical vapor deposition (CVD) of TiX
and of the kinetics in the formation of excess layers.
\textit{The upper panel} gives a principle sketch
of a CVD apparatus.
A supply gas mixture that contains several different gases with 
different concentrations $c_i$ is injected 
into the CVD chamber via valve 1 at a rate $R_{\text{sup}}$.
The chamber is kept at a constant temperature $T$
and pressure $p$. 
Inside the chamber the gases react and form TiX.
Reaction products and unused supply gas
are exhausted at a rate $R_{\text{exh}}$ via valve 2.
\textit{The set of lower panels} illustrates  
kinetic effects that specify the surface termination.
A Ti (X) surface layer can be deposited
from a Ti (X) carrying gas,
here shown for TiCl$_4$.
As indicated, the process involves a set of unknown complex
transition states (TS).
}
\end{figure}

\subsection{Materials example: TiC \& TiN\label{sec2}}
TiX(111) (X = C or N) are important, specific, examples
where surface terminations have industrial relevance.
Specifically, alternating deposition of 
TiX and \alumina\ by CVD
\cite{Ruppi200150,Larsson2002203}
is commonly used to produce
wear-resistant coatings on cemented-carbide cutting tools.
\cite{Halvarsson1993177}

The top panel of figure \ref{fig:CVDschematics} illustrates
the CVD process used for TiX growth.
A TiCl$_4$-CH$_4$-H$_2$-HCl (for TiC \cite{Halvarsson1993177}) 
or TiCl$_4$-N$_2$-H$_2$ (for TiN \cite{Larsson2002203}) supply gas mixture
is injected into a CVD chamber via valve 1.
The chamber is kept at a high constant temperature
and at a constant pressure. 
The overall reactions that lead to the deposition of
TiX are
\begin{subequations}
\label{eq:CVDTiX}
\begin{align}
\text{TiCl$_4$}+\text{CH$_4$}&
\rightarrow
\text{TiC}+4\text{HCl},
\label{eq:CVDTiC}\\
\text{TiCl$_4$}+\frac{1}{2}\text{N$_2$}+2\text{H$_2$}&
\rightarrow
\text{TiN}+4\text{HCl}.
\label{eq:CVDTiN}
\end{align}
\end{subequations}
Unused supply gas as well as reaction products
are exhausted from the chamber via valve 2.

In TiX/alumina multilayers, 
the TiX(111) surface terminations 
play a crucial role for the nucleation of alumina 
and consequently for the nature 
of the TiX/alumina interface in the coatings.
In particular, Ti-terminated surfaces indicate 
an interface composition of the type 
TiX:Ti/O:\alumina,
whereas X-terminated surfaces favor
a  TiX:X/Al:\alumina\ interface.
\cite{ruberto:235438,Vojvodic20063619,VojvodicLiC}
Here, (O,Al):\alumina\ specifies the dominant
chemical species (O or Al) of alumina
at the interface. 

To the best of our knowledge,
experimental studies on the TiX(111) surface termination 
exist only for TiC, \cite{JJAP.20.L829,Oshima198169,Zaima1985380}
but not for TiN.
They all report a Ti-terminated surface.
However, in all these experiments, the TiC(111) surface
was prepared by high-temperature annealing
(and associated selective evaporation) under UHV conditions.
The observed Ti-termination of the TiC(111) surface
under such experimental conditions
can be understood from a theoretical argument due to 
\textit{Tan} \etal,\cite{Tan199649}
estimating the selective high-temperature evaporation
from the total-energy cost of pulling
atoms out of material.

A predictive, atomistic  theory of the surface termination 
of as-deposited TiX(111)
as a function of the growth environment 
remains of central value.
The characterizations and predictions are important,
for example, 
to define a proper starting point for
understanding atomistic processes
in the  fabrication of TiX/alumina multilayer coatings.
\cite{Ruppi200150}

\subsection{Kinetics of deposition processes}
The bottom panel of Fig.~\ref{fig:CVDschematics} shows a schematics 
of the kinetics leading to a surface system with an excess Ti layer.
The initial state is represented by a TiCl$_4$ molecule
above the X-terminated TiX(111) surface.
In the final state, a Ti atom (or a layer after
a number of events) is deposited.
Additional molecules that enter or leave
the reaction and produce the final state
are not shown.
We also sketch the energy landscape
as a function of a generic reaction coordinate.
We note that only the relative energy positions of
the initial and final states are shown in correspondence
with the actual results of  our modeling;
the nature and energy of the relevant transition states are unknown.

Surface morphologies obtained in MBE growth can often by understood 
from kinetic Monte Carlo (kMC) simulations,
using DFT calculations to determine process barriers.\cite{kMC_Kratzer,kMC_Bogicevic,PhysRevB.72.115401,PhysRevLett.83.2608}
Such simulations follow the surface morphology over some time
by allowing for a number of events to take place at random.
Conceptually it is simple to generalize the kMC approach to CVD, 
but in practice such a generalization will be extremely challenging.
The chemical species supplied in CVD are usually more
complex than single atoms or dimers used in MBE.
Worse, the reactants will not, in general, simply dissociate
on the surface but enter into complicated reactions
with other species and form clusters.
These reactions are not limited to the surface
and may lead to a large variety of intermediates.
While adaptive intelligent kMC exists and is being developed,\cite{PhysRevB.72.115401}
the complexity in CVD is enormous.

At the same time,  the high deposition temperature in CVD 
motivates a shift away from explicit kinetic modeling.
We therefore develop a thermodynamic description
that focuses entirely on initial and final states,
retaining the simplicity of the analysis
for selective evaporation in UHV. \cite{Tan199649}
A nonequilibrium thermodynamic description becomes applicable
when high temperature facilitates structural reorganization and permits
the growing system to reach and sample the set of possible final states.

\subsection{Excess-layer deposition}
\begin{figure}
\begin{tabular}{c}
\epsfig{file=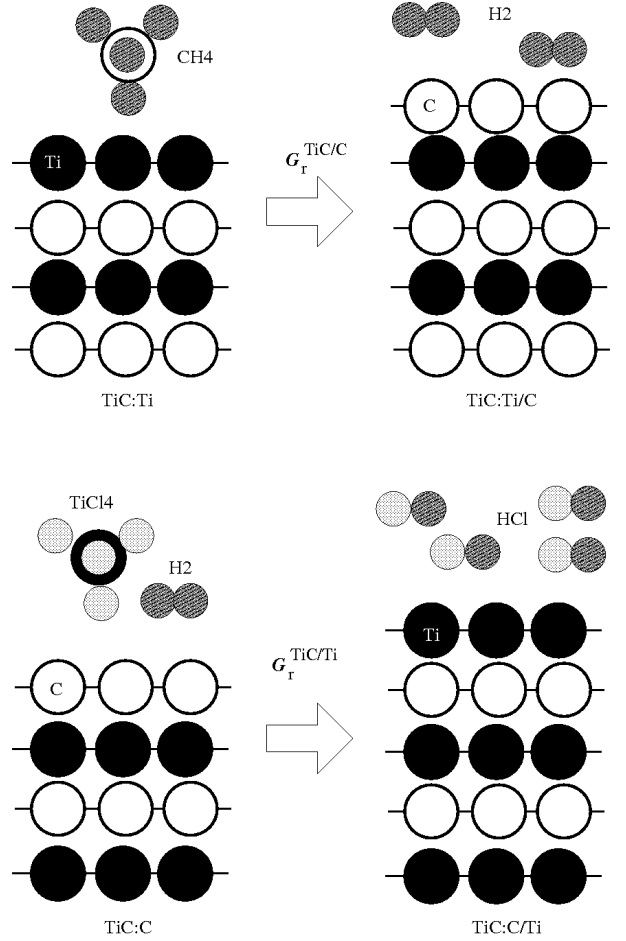,width=8cm}
\end{tabular}
\caption{
\label{fig:Hnuc}
Formation of excess layers and associated 
free energies of reaction $G_{\text{r}}$.
The top panel illustrates the formation 
of a C excess layer from CH$_4$
on a Ti-terminated surface.
The bottom panel shows the formation of a Ti excess layer
from TiCl$_4$ and H$_2$
on a C-terminated surface.
From a kinetic perspective,
formation of specific clusters and presence of
H$_2$ may catalyze the process shown in
the upper panel;
thermodynamically, such catalytic processes
do not, however,  affect the free energy of reaction.
}
\end{figure}

Figure \ref{fig:Hnuc} illustrates possible processes 
that lead to deposition of 
single Ti and C layers for TiC.
We determine a preference for the surface termination
by comparison of the Gibbs free energy of reaction
for such excess-layer structures
with those of the stoichiometric TiX.
The reactions listed in (\ref{eq:CVDTiX}) only describe 
deposition that leads to TiX with 
the full stoichiometry of the bulk.
We supplement that description by a thermodynamic characterization
of dominant reaction pathways leading to formation 
of excess layers for both TiC and TiN systems.

For excess-layer reactions,
we consider the following reactions for Ti deposition
on TiC and TiN,
\begin{subequations}
\label{eq:excessTi}
\begin{align}
\label{eq:TiC/Ti}
&\text{TiC:C}+
\text{TiCl$_4$}+2\text{H$_2$}
\rightarrow
\text{TiC:C/Ti}+4\text{HCl},\\
\label{eq:TiN/Ti}
&\text{TiN:N}+
\text{TiCl$_4$}+2\text{H$_2$}
\rightarrow
\text{TiN:N/Ti}+4\text{HCl}.
\end{align}
\end{subequations}
Here, TiC:C/Ti (TiN:N/Ti) identifies a system with
an additional Ti layer on the C-terminated TiC 
(N-terminated TiN) surface.
Additional X layers may be deposited as
\begin{subequations}
\label{eq:excessX}
\begin{align}
\label{eq:TiC/C}
&\text{TiC:Ti}+
\text{CH$_4$}
\rightarrow
\text{TiC:Ti/C}+2\text{H$_2$},\\
\label{eq:TiN/N}
&
\text{TiN:Ti}+\frac{1}{2}\text{N$_2$}
\rightarrow \text{TiN:Ti/N}.
\end{align}
\end{subequations}
Since deposition of Ti is relevant only on an
X-terminated surface and deposition of X only 
on the Ti-terminated surface
we suppress the specification
of the original surface termination
in the following.
For example, TiC/Ti will be used instead
of  TiC:C/Ti.

\section{\textit{Ab-initio} thermodynamics of deposition growth \label{sec:AIT-DG}}

Our thermodynamic method to predict terminations of growing surfaces
describes phenomena that lie between the static equilibrium limit
and the non-equilibrium limit
that requires a fully kinetic theory.
We assume an experimental setup similar
to that illustrated in the top panel of 
figure~\ref{fig:CVDschematics}.

\subsection{Computational strategy and assumptions\label{sec:Strategy}}
We use the difference in Gibbs free energies of reaction
\begin{align}
\Delta G_r^{\text{AB}}=G_r^{\text{A}}-G_r^{\text{B}} 
\label{eq:predictor}
\end{align}
as a predictor for the prevalence of growing either an A-terminated ($\Delta G_r^{\text{AB}}<0$)
or a B-terminated  ($\Delta G_r^{\text{AB}}>0$) surface of a binary material AB.
Here, $G_r^{\text{A(B)}}$ is the free energy of reaction associated
with the formation of an excess A (B) layer.
Our choice of predictor follows naturally from
an analogy with chemical reaction theory,\cite{Demirel,Tschoegl}
and from considering surface transformations.
The following coupled set of reactions 
change an A- (B-) terminated AB surface
into a B- (A-) terminated surface,
\begin{subequations}
\begin{align}
\label{eq:R1}
I:~\text{B}+\text{React}_I
\mathop{\rightleftharpoons}^{\Gamma_f^I}_{\Gamma_b^I} 
\text{A}+\text{Prod}_I\\
\label{eq:R2}
II:~\text{A}+\text{React}_{II}
\mathop{\rightleftharpoons}^{\Gamma_f^{II}}_{\Gamma_b^{II}} 
\text{B}+\text{Prod}_{II}.
\end{align}
\label{eq:reactions}
\end{subequations}
The probability for growing either an A- or B-terminated surface 
is described by rate equations
where the rates $\Gamma_{\{f,b\}}^{\{I,II\}}$
depend on the deposition environment
(reactants $\text{React}_{\{I,II\}}$ and products $\text{Prod}_{\{I,II\}}$).
The steady-state solution $P_{\text{A}}/P_{\text{B}}=
(\Gamma_f^{I}+\Gamma_b^{II})/
(\Gamma_b^{I}+\Gamma_f^{II})$
is closely approximeted by
\begin{align}
P_{\text{A}}/P_{\text{B}}\approx\exp\left(-\beta\Delta G_r^{AB}/2\right),
\label{evaluation}
\end{align}
where $\beta$ denotes the inverse temperature.
This evaluation (\ref{evaluation}) becomes exact in the limit of dynamic equilibrium
(identified by $G_r^{\text{A}}+G_r^{\text{B}}=0$)
where $\Delta G_r^{\text{AB}}$ reduces to a difference in surface free energies.
Further details on the nature of the strategy and the applicability
to nonequilibrium conditions are given in the appendix.

The assumptions that facilitate the evaluation
of free energies of reaction are:
\begin{enumerate}
\item The system composed of the
growing surface and the gas environment 
settles into a steady state,
specified by the gas supply,
deposition, and gas exhaust rates.
We do not \textit{a priory} assume
that the gases and the surface are
in mutual equilibrium,
nor do we generally expect such an equilibrium to emerge.
\item Each of the gases that take part
in reactions is fully thermalized 
at the steady-state partial pressure
so that we can associate a chemical potential
to them.
\item The overall pressure is low enough
so that a description of all gases
in terms of the ideal-gas approximation
applies.
\item The temperature of the system is high enough
so that we can neglect kinetic barriers.
\end{enumerate}

The computational strategy 
and the first assumption of a steady state is a novel feature
of our method.
We derive the steady state from a rate equation
that describes the supply and exhaust of gases
into and from the reaction chamber
as well as the deposition inside the chamber.

The other three assumptions are identical to those
of the AIT-SE method.
The AIT-SE describes a system in a static equilibrium
at high temperatures.
The AIT-SE might also describe systems in \textit{dynamic equilibrium}.
Dynamic equilibrium\cite{Tschoegl} here means
that the free energy of reaction (derived and calculated below)
exactly equals zero.
This condition also applies in static equilibrium
but in dynamic equilibrium the zero-value of the free energy of reaction
arises from the steady-state concentrations of reactants and reaction products.
A (virtual) drop in the reaction rate 
readily leads to a drop of the concentrations of reaction products
but not of the reactants (due to the supply and exhaust),
so that the dynamics of the reaction is restored.

We find that as the steady state approaches dynamic equilibrium,  
the predictions of AIT-DG 
coincide with those of the AIT-SE.
However, even if dynamic equilibrium is reached,
our description in terms of a rate equation
will be necessary in order to determine  chemical potentials
of the gas-phase constituents
from their individual pressures
during the process.
We also emphasize that, during growth,
the steady state does not necessarily
settle into a state of dynamic equilibrium.

\subsection{Excess layers and their free energies of reaction}
Figure \ref{fig:Hnuc} illustrates the formation
of Ti and C excess layers on TiC(111),
corresponding to the reactions in (\ref{eq:TiC/Ti}) and (\ref{eq:TiC/C}).
Similarly, we assume that Ti and N excess layers
can be deposited on TiN(111) according to
the  reactions in (\ref{eq:TiN/Ti}) and (\ref{eq:TiN/N}).

We represent the TiX(111) 
surface before deposition of additional layers 
by a stoichiometric slab.
Upon deposition of an additional layer
the slab becomes non-stoichiometric, 
which motivates the terminology of an excess layer.
A stoichiometric TiX(111) slab is asymmetric along the [111] direction.
In particular, it possesses one Ti-terminated surface
and one X-terminated surface.
An excess  Ti  layer is deposited on the X-terminated
side of the slab,
while an excess X  layer is be deposited 
on the Ti-terminated side.

We introduce the chemical potentials $\mu_{\text{Ti}}^{\text{aig}}$,
$\mu_{\text{C}}^{\text{aig}}$, and $\mu_{\text{N}}^{\text{aig}}$
for Ti, C, and N atoms in the gases.
The set of $\mu^{\text{aig}}_i$ can be conveniently
related to differences between 
several molecular chemical potentials
(see below).
The chemical potentials
of Ti, C, and N in the gases  do not in general equal  
the respective chemical potentials 
$\mu_{\text{Ti}}$,
$\mu_{\text{C}}$, and $\mu_{\text{N}}$
in the solid bulk.
For the latter the free energy per bulk unit,\cite{bulkFreeEnergy}
$g_{\text{TiX}}$,
always specifies material stability,
\begin{align}
\mu_{\text{Ti}}+\mu_{\text{X}} = g_{\text{TiX}}.
\label{eq:SurfaceBulkEquilibrium}
\end{align}
Growth of TiX, on the other hand, is characterized by
\begin{align}
g_{\text{TiX}}<
\mu_{\text{Ti}}^{\text{aig}}+\mu_{\text{X}}^{\text{aig}}.
\label{eq:GrowthCondition}
\end{align}
The free energy of reaction\cite{Demirel} for the deposition of one unit of bulk 
is defined as
\begin{align}
g_{\text{r}}^{\text{TiX}}=g_{\text{TiX}}-
\mu_{\text{Ti}}^{\text{aig}}-\mu_{\text{X}}^{\text{aig}}.
\label{eq:GrBulk}
\end{align}

We also introduce the free energy of reaction $G_{\text{r}}$
which corresponds to the gain in energy per surface unit
in the deposition of one excess layer.
For deposition of excess Ti, $G_{\text{r}}$ can then be written as
\cite{EquivalentDefinition}
\label{eq:GrTiX/Ti}
\begin{subequations}
\begin{align}
G_{\text{r}}^{\text{TiC/Ti}}&
=G_{\text{TiC/Ti}}-G_{\text{TiC}}-\mu_{\text{Ti}}^{\text{aig}}
\label{eq:GrTiC/Ti}\\
G_{\text{r}}^{\text{TiN/Ti}}&
=G_{\text{TiN/Ti}}-G_{\text{TiN}}-\mu_{\text{Ti}}^{\text{aig}}.
\label{eq:GrTiN/Ti}
\end{align}
\end{subequations}
Here, $G_{\text{TiX}}$ is the free energy of a
stoichiometric, non-symmetric TiX slab
and $G_{\text{TiX/Ti}}$
denotes the free energy of the system
that consists of the same stoichiometric TiX slab 
and an additional Ti excess layer 
(adsorbed on the X-terminated side of the slab).
For excess X, the free energy of reaction is given by
\begin{subequations}
\begin{align}
G_{\text{r}}^{\text{TiC/C}}&=G_{\text{TiC/C}}
-G_{\text{TiC}}-\mu_{\text{C}}^{\text{aig}}
\label{eq:GrTiC/C}\\
G_{\text{r}}^{\text{TiN/N}}&
=G_{\text{TiN/N}}-G_{\text{TiN}}-\mu_{\text{N}}^{\text{aig}}.
\label{eq:GrTiN/N}
\end{align}
\label{eq:GrTiX/X}
\end{subequations}

With these definitions, a negative value of  $G_{\text{r}}$
implies that the deposition
of the excess layer is thermodynamically favorable
and that the reaction proceeds in the direction
of the arrow in one of the listings
(\ref{eq:excessTi}) and (\ref{eq:excessX}).
A positive value of $G_{\text{r}}$ implies 
that the deposition of the
corresponding excess layer is thermodynamically unfavorable
and that the reaction proceeds in the opposite direction.

The atom-in-gas chemical potentials $\mu_{i}^{\text{aig}}$
depend on the composition of the environment
and can therefore be controled by the gas flow.
For the case of CVD of TiX with the here-considered
supply gas and the reactions listed in
(\ref{eq:excessTi}) and (\ref{eq:excessX}),
we assign the atomic chemical potentials
in the gas as
\begin{subequations}
\label{eq:muAIG}
\begin{align}
\mu_{\text{Ti}}^{\text{aig}}&=\mu_{\text{TiCl$_4$}}+2\mu_{\text{H$_2$}}-4\mu_{\text{HCl}}\\
\mu_{\text{C}}^{\text{aig}}&=\mu_{\text{CH$_4$}}-2\mu_{\text{H$_2$}}\\
\mu_{\text{N}}^{\text{aig}}&=\frac{1}{2}\mu_{\text{N$_2$}}.
\end{align}
\end{subequations}

We note that,  in general,
the deposition of excess Ti 
followed by the deposition of excess X (or vice versa)
equals the free energy of reaction 
for the deposition of one unit of bulk,
\begin{align}
G_{\text{r}}^{\text{TiX/Ti}}+G_{\text{r}}^{\text{TiX/X}}=g_{\text{r}}^{\text{TiX}}.
\label{eq:GrBulk}
\end{align}
If the solid-gas system approaches \textit{equilibrium} 
(mediated through the surface),
we have $g_{\text{r}}^{\text{TiX}}=0$,
so that the reaction energy gain is simply
\begin{align}
G_{\text{r}}^{\text{TiX/Ti}}+G_{\text{r}}^{\text{TiX/X}}=0.
\label{eq:GrEquilibrium}
\end{align}

We emphasize that the definitions (\ref{eq:GrTiX/Ti}) and  (\ref{eq:GrTiX/X}) 
do not make any assumption about equilibrium.
On the contrary, by calculating (\ref{eq:GrBulk})
as a function of deposition pressure and temperature,
we can predict whether the process is in or out of equilibrium,
and in which direction it proceeds.

\subsection{Gibbs free energies and chemical potentials. }
Calculating free energies of reactions requires
simple approximations to evaluate free energies
of surface systems and  chemical potentials
of gas-phase species.
We base these approximations on
experimentally available thermochemical data.

For the gases, we employ the ideal gas approximation.
This approximation in justified for low pressures
which is typically the case in CVD of TiX.
It also allows us to express the $\mu_i$ 
in terms of temperature $T$ and partial pressure $p_i$,
\begin{align}
\mu_i(T,p_i)=\epsilon_i+\Delta_i^0(T)+k_BT\ln(p_i/p^0).
\label{eq:mu}
\end{align}
Here $\epsilon_i$ is the DFT total energy 
of the gas phase species (molecule),
and $k_B$ is  the Boltzmann constant.
$\Delta_i^0(T)$ is the temperature dependence
of $\mu_i$ at a fixed pressure $p^0$,
and related to enthalpy and entropy differences. \cite{AIT_Scheffler}
The latter are available for $p^0=1$~atm 
for many molecules in thermochemical tables such as
Ref.~\onlinecite{JANAF}.

For surface systems, we approximate the Gibbs free energy
by the DFT total energy,
\begin{align}
G\approx E,
\label{eq:approxG}
\end{align}
where $G$ and $E$ stand for one
of $G_{\text{TiX/Ti}}$ and $E_{\text{TiX/Ti}}$
or $G_{\text{TiX/X}}$ and $E_{\text{TiX/X}}$.
This approximation neglects vibrational contributions
($TS^{\text{vib}}$) and the pressure term
($pV$) in $G$.
However, the effect of neglecting these factors
is expected to be small.\cite{AIT_Scheffler,SurfaceVibComment}

\subsection{Rate equation and steady-state solution}
Evaluation of the gas-phase chemical potentials
(\ref{eq:mu})
requires the specification of the deposition temperature 
and the individual partial pressures.
The temperature and the total pressure inside
the CVD chamber can be  controlled reasonably well
during fabrication.
Individual partial pressures,
on the other hand,
are not directly accessable.
However, they can be determined 
from a rate equation 
that describes the overall CVD process.

Using the ideal gas approximation ($pV=Nk_BT$),
the change in individual pressures is described by
\begin{align}
\partial_t p_i(t)=\frac{k_B T}{V_{\text{CVD}}}\left[ c_iR_{\text{sup}}-\frac{p_i(t)}{p(t)}R_{\text{exh}}+
\nu_iR_{\text{TiX}}\right],
\label{eq:PressureEvolution}
\end{align}
where $p(t)=\sum_i p_i(t)$ is the total pressure,
$p_i(t)$ denotes the partial pressure 
of species $i$ at time $t$,
and $\nu_i$ is the corresponding stoichiometric coefficient
(change in number of molecules) in the deposition process,
Eq.~(\ref{eq:CVDTiX}).
Per definition, \cite{Demirel}
stoichiometric coefficients 
are negative for species that are consumed in the
reaction
and positive for reaction products.
The value of $c_i$ reflects the concentration 
of species $i$ in the supply gas.
Finally, $R_{\text{sup}}$ and $R_{\text{exh}}$
are the rates at which the gases are 
supplied to and exhausted from the chamber;
$R_{\text{TiX}}$ is the rate  of TiX deposition
(per stoichiometric formula).

In steady state, we can eliminate either
one of the three rates.
Keeping the supply and deposition rate as fundamental variables,
the steady-state solution ($\partial_tp=0$ and $\partial_tp_i=0$) 
of (\ref{eq:PressureEvolution}) is given by 
\begin{align}
p_i&=p\,\frac{c_i+\nu_i\,r_{\text{TiX}}}{1+\Delta\nu\,r_{\text{TiX}}}.
\label{eq:SSpressures}
\end{align}
Here, $p$ and $p_i$ (without time argument)
denote the steady-state total (which can be controlled)
and partial pressures
and $\Delta\nu=\sum\nu_i$
is the number of molecules that are created
in each reaction.
We  have also introduced the scaled TiX deposition rate 
$r_{\text{TiX}}=R_{\text{TiX}}/R_{\text{sup}}$.

Collecting equations
(\ref{eq:GrTiX/Ti}),
(\ref{eq:GrTiX/X}),
(\ref{eq:mu}),
(\ref{eq:approxG}),
and
(\ref{eq:SSpressures}),
we can express 
the Gibbs free energy of reaction
for deposition of excess layers 
as a function of three variables:
the temperature $T$, the pressure $p$
and the scaled TiX deposition rate $r_{\text{TiX}}$
\begin{align}
G_{\text{r}}=
G_{\text{r}}\left(\{\mu_i(T,p_i(r_{\text{TiX}}))\}\right)
=G_{\text{r}}(T,p,r).
\end{align}
Here $G_{\text{r}}$ denotes one of
$G_{\text{r}}^{\text{TiX/Ti}}$
or $G_{\text{r}}^{\text{TiX/X}}$.

\subsection{Limits on the scaled deposition rate}
The scaled deposition rates $r_{\text{TiC}}$ 
and $r_{\text{TiN}}$ 
possess critical values $r^{\text{crit}}_{\text{TiC}}$
and $r^{\text{crit}}_{\text{TiN}}$,
\textit{i.e.}, upper values for the parameter $r$
that are still compatible with actual growth.
Deposition of TiX is favorable only
if the inequality (\ref{eq:GrowthCondition}) holds.
For $r_{\text{TiX}}\ge r^{\text{crit}}_{\text{TiX}}$, 
the accumulation of reaction products
makes the reaction thermodynamically unfavorable.
Thus, in a steady state, 
$r_{\text{TiX}}\le r^{\text{crit}}_{\text{TiX}}$ must be fulfilled
whereas equality corresponds to (dynamic) equilibrium.
This leads to the conditions
\begin{subequations}
\label{eq:rCrit}
\begin{align}
g_{\text{TiC}}\equiv&
\mu_{\text{TiCl$_4$}}(r^{\text{crit}}_{\text{TiC}})+
\mu_{\text{CH$_4$}}(r^{\text{crit}}_{\text{TiC}})
-+4\mu_{\text{HCl}}(r^{\text{crit}}_{\text{TiC}}),\\
g_{\text{TiN}}\equiv&
\mu_{\text{TiCl$_4$}}(r^{\text{crit}}_{\text{TiN}})
+2\mu_{\text{H$_2$}}(r^{\text{crit}}_{\text{TiN}})
+\frac{1}{2}\mu_{\text{N$_2$}}(r^{\text{crit}}_{\text{TiN}})\nn\\
&-4\mu_{\text{HCl}}(r^{\text{crit}}_{\text{TiN}}),
\end{align}
\end{subequations}
using  $\mu_{i}(r^{\text{crit}}_{\text{TiX}})=\mu_{i}(T,p[r^{\text{crit}}_{\text{TiX}}])$ 
as shorthand.
Critical values, \textit{i.e.}, upper limits for the scaled deposition rate, 
are identified by solving these equations.

\subsection{Equilibrium limit\label{sec:EquilibriumLimit}}
The AIT-DG method formally resembles the AIT-SE 
when taking the limit in which the gas environment
is in equilibrium with the surface.
Strictly speaking, the AIT-DG method 
expresses the same thermodynamic preference as 
does a generalization of the AIT-SE method to  dynamic equilibrium.
The dynamic equilibrium is nevertheless
still specified  by solving the rate equations 
for the steady-state gas flow.

We express the differences in reaction free energies 
for excess Ti and excess X surface layers as
\begin{align}
G_{\text{r}}^{\text{TiX/Ti}}-G_{\text{r}}^{\text{TiX/X}}=&
G_{\text{TiX/Ti}}-\mu_{\text{Ti}}^{\text{aig}}\nn\\
&-\left(G_{\text{TiX/X}}-\mu_{\text{X}}^{\text{aig}}\right).
\end{align}

Equilibrium between the gas phase and the surface,
Eq.~(\ref{eq:GrEquilibrium}), implies
\begin{align}
0&
=\mu_{\text{Ti}}^{\text{surf}}+\mu_{\text{X}}^{\text{surf}}
-\mu_{\text{Ti}}^{\text{aig}}-\mu_{\text{X}}^{\text{aig}},
\end{align}
where $\mu_{\text{Ti}}^{\text{surf}}=G_{\text{TiX/Ti}}-G_{\text{TiX}}$
and $\mu_{\text{X}}^{\text{surf}}=G_{\text{TiX/X}}-G_{\text{TiX}}$
are the chemical potentials of Ti and X at the
surface of the solid.

If we also assume equilibrium between the bulk and the surface,
that is, 
$\mu_{\text{Ti}}^{\text{surf}}+\mu_{\text{X}}^{\text{surf}}
=\mu_{\text{Ti}}+\mu_{\text{X}}$,
and use (\ref{eq:SurfaceBulkEquilibrium}),
we obtain
\begin{align}
G_{\text{r}}^{\text{Ti}}-G_{\text{r}}^{\text{X}}&=
G_{\text{TiX/Ti}}-G_{\text{TiX/X}}
-g_{\text{TiX}}+2\mu_{\text{X}}^{\text{aig}}
\label{eq:equilibriumLimit}
\end{align}
as a predictor of preference
of \textit{equilibrium} surface terminations.

This equilibrium predictor can be expressed in terms of
a difference in surface free energies,
thus making a connection to the AIT-SE,\cite{O2vsN2}
\begin{align}
G_{\text{r}}^{\text{Ti}}-G_{\text{r}}^{\text{X}}=
2A\left(\gamma^{\text{TiX/Ti}}-\gamma^{\text{TiX/X}}\right).
\end{align}
Here, we  use the standard definition of the
surface free energy 
\begin{subequations}
\begin{align}
\gamma^{\text{TiX/Ti}}&=\frac{1}{2A}\left(
G_{\text{TiX/Ti}}
-n_1g_{\text{TiX}}+\mu_{\text{X}}
\right),\\
\gamma^{\text{TiX/X}}&=\frac{1}{2A}\left(
G_{\text{TiX/X}}
-n_2g_{\text{TiX}}-\mu_{\text{X}}
\right),
\end{align}
\end{subequations}
for the Ti- and X-terminated surfaces, respectively.
The numbers  $n_1$ and $n_2$ correspond to the numbers of 
Ti layers in the TiX/Ti and the TiX/X slabs,
so that  $n_1-n_2=1$.
In equilibrium, the difference in free energies of reaction 
is therefore simply the difference in surface free energies
between the Ti- and X-terminated surface

At the same time, we emphasize that,
\begin{align}
\mu_{\text{Ti}}+\mu_{\text{X}}=g_{\text{TiX}}
\neq\mu_{\text{Ti}}^{\text{aig}}+\mu_{\text{X}}^{\text{aig}}
\end{align}
applies for general growth conditions.
A nonequilibrium-driven process like growth
will not generally adjust itself to dynamic equilibrium.
Out of dynamic equilibrium, a prediction of surface terminations 
in terms of surface free energies 
(employing  $\mu_{\text{X}}=\mu_{\text{X}}^{\text{aig}}$)
is not justified.
Instead, the free energy of reaction
is the proper quantity to use
to predict surface terminations.

\section{Computational Method\label{sec4}}
All DFT calculations are performed with the plane-wave code 
\textsf{dacapo} \cite{ref:DACAPO}
using ultra-soft pseudopotentials, \cite{PhysRevB.41.7892}
the PW91 exchange-correlation functional, \cite{PhysRevB.46.6671}
and 400~eV  plane-wave cutoff.

The 1$\times$1 TiX(111) surfaces are represented by  slab geometry
within a supercell including $\sim10$~\AA\ of vacuum.
A $3\times3\times1$ Monkhorst-Pack  k-point sampling\cite{PhysRevB.13.5188} is used
and atomic relaxations are performed until interatomic forces
no longer exceed a value of 0.03~eV/\AA.
The slabs contain 12 Ti and 12 X layers for the stoichiometric
reference surface without excess layer,
13 Ti and 12 X layers for the surface with Ti excess layer,
and 12 Ti and 13 X layers for the surface with  X excess layer.
We have checked that the bulk energies per TiC unit, 
calculated as
$\epsilon_{\text{TiX}}=E(N+1)-E(N)$,
are converged to a difference of less than 0.03~eV
with respect to the slab thickness.
Here $E(N)$ and $E(N+1)$
is the energy of a TiX, TiX/Ti, or TiX/X slab
that contains $N$ and $N+1$ Ti layers,
that is, the bulk energy
is independent of the surface termination,
as it should be.

For molecules and atoms, we use spin-polarized
calculations when relevant.
Total energies are calculated within
unit cells of size $20\times20\times20$~\AA$^3$,
assuring that molecules from different cells do not interact.
We use a $1\times1\times1$ Monkhorst-Pack k-point sampling\cite{PhysRevB.13.5188}
and perform atomic relaxations  until interatomic forces
no longer exceed a value of 0.01~eV/\AA.

\begin{table*}
\begin{ruledtabular}
\begin{tabular}{ccccccccc}
molecule  &   \mc{2}{c}{$E_{\text{atom}}$ [eV/molecule]}& $\Delta E_{\text{atom}}$ [\%]  
&   \mc{2}{c}{$b$ [\AA]}  & $\Delta b$ [\%]    	& \mc{2}{c}{$\theta$ [$^{\circ}$]}\\
          &   present  & exper.\footnotemark[1] &       & present & exper.\footnotemark[1] & &present & exper.\footnotemark[1]\\
\hline
H$_2$	  & 4.58  & 4.48  & 2.2\%  &  0.75    & 0.741 & 1.2\%	&  ---    & ---  \\ 
TiCl$_4$  &21.31  & 17.84 & 19\%   &  2.18    &  2.17 & 0.05\%   & 109.3-7 &  109.471 \\
N$_2$	  & 9.63  & 9.76  & 1.3\%  &  1.12    & 1.098 & 2.0\%	&  ---    & ---  \\ 
CH$_4$	  &18.23  & 17.02 & 7\%    &  1.10    & 1.087 & 1.2\%    & 109.5   &  109.471 \\
HCl	  & 4.88  & 4.43  & 10\%   &  1.29    & 1.275 & 1.2\%	&  ---    &    ---  
\end{tabular}
\end{ruledtabular}
\footnotetext[1]{Ref.~\onlinecite{NISTweb}} 
\caption{
\label{tab:molecules}
Calculated and experimental data for molecules relevant to CVD of TiX.
Total energies are required for reference for the calculation of chemical potentials.
Calculated geometric data, such as bond lengths $b$ and bond angles $\theta$, 
is in excellent agreement with the experimental one.
Calculated atomization energies $E_{\text{atom}}$ 
of N$_2$, H$_2$, and CH$_4$ differ only slightly from experimental values
(We use the value of $E_{\text{H}}=-13.6$~eV for the H atom
in favor of the DFT value to calculate atomization energies of molecules
that carry H).
For Cl-containing molecules, the differences are larger.
}
\end{table*}

\begin{figure}
\epsfig{file=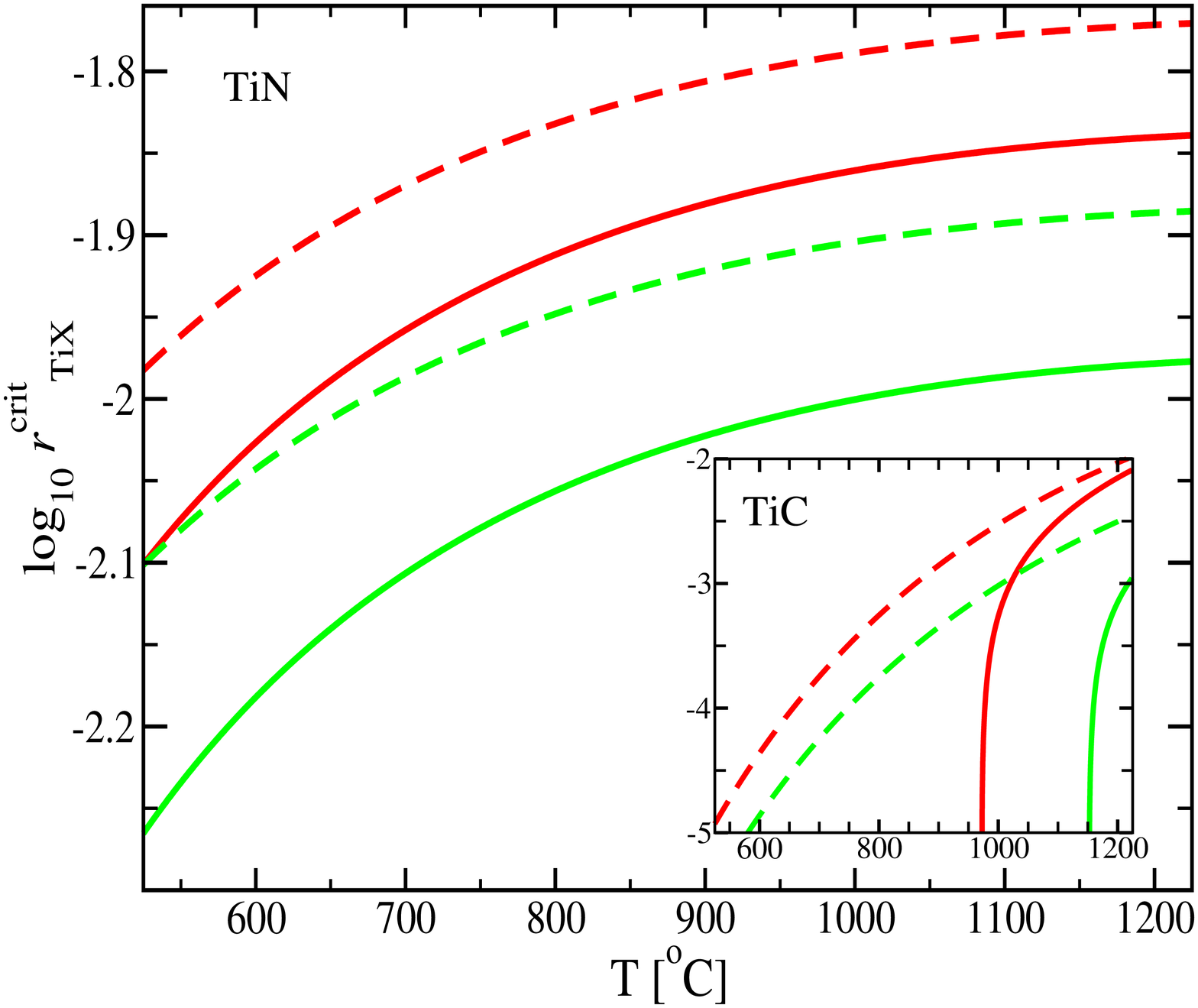,width=8cm}
\caption{
\label{fig:rCrit}
(Color online)
Critical values of the scaled reaction rate
$r^{\text{crit}}_{\text{TiN}}$ as functions of
temperature at different pressures and
HCl concentrations.
Red (dark) lines correspond to a 
deposition pressure of $p_1=50$~mbar,
green (light) lines to $p_2=500$~mbar.
Solid lines correspond to no HCl content $c_{\text{HCl},1}=0$ 
in the supply gas,
dashed lines to a content of 1\% HCl
$c_{\text{HCl},1}=0.01$.
The insert shows the corresponding analysis for TiC.
}
\end{figure}

\begin{figure}
\begin{tabular}{c}
\epsfig{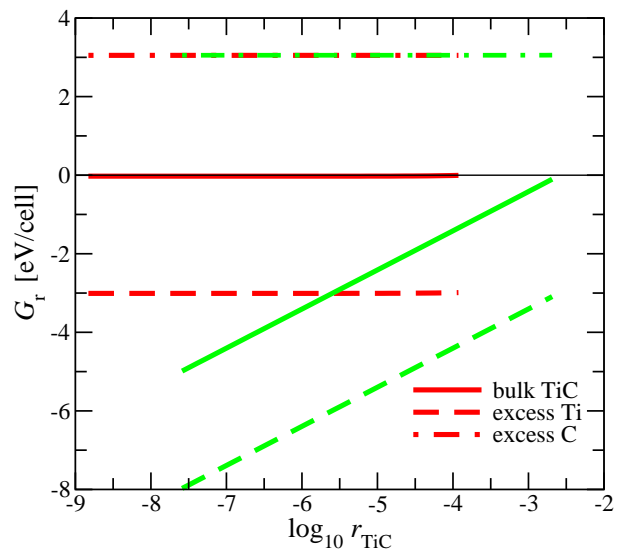}
\end{tabular}
\caption{
\label{fig:HnucCVD}
Free energies of reaction $G_{\text{r}}$ of CVD 
bulk TiC (solid), excess Ti (dashed), and C excess layers (dashed-dotted) 
at the TiX(111) surface 
as a function of the scaled reaction rate
$r_{\text{TiC}}$.
We assume 
a total pressure of $p=50$~mbar
and a deposition temperature of $T=980^{\circ}$C
inside the CVD chamber.
Red (dark) lines correspond to
experimental values\cite{Halvarsson1993177} 
for the supply gas compositions,
see also main text.
In particular, the supply gas contains 
1\% HCl.
The green (light) lines correspond to 
the case where no HCl is supplied.
}
\end{figure}

\begin{figure}
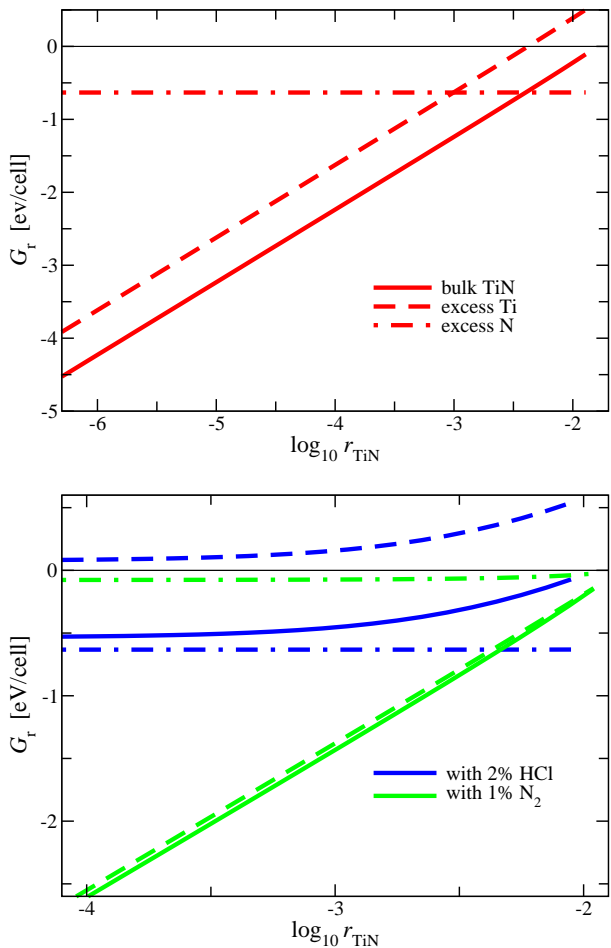

\begin{tabular}{c}
\epsfig{file = fig6a.eps, width=8.0cm}\\[0.3cm]
\epsfig{file = fig6b.eps, width=8.0cm}
\end{tabular}
\caption{
\label{fig:HnucCVDTiN}
Free energies of reaction $G_{\text{r}}$ of CVD 
bulk TiN, excess Ti and excess N layers at the TiX(111) surface 
as a function of the scaled reaction rate
$r_{\text{TiN}}$.
In the top panel,
experimental values \cite{Larsson2002203} 
for the supply gas compositions
from  are used,
see also main text.
Temperature and pressure are chosen as in 
figure~\ref{fig:HnucCVD}.
The termination critically depends on the 
the scaled deposition rate.
The bottom panel illustrates the effect of varying
deposition parameters.
For bulk, excess Ti, and excess N,
the same line style as in the upper panel applies.
Blue (dark) lines correspond to an HCl concentration of 2\%
in the supply gas.
Green (light) lines correspond to the case where 
the N$_2$ concentration is decreased to 1\%
and the temperature is raised to $T=1200^{\circ}$C.
In both cases, the changes are assumed to be balanced
by the H$_2$ concentration.
An increased HCl concentration leads to a N-terminated surface,
independently the value  of $r_{\text{TiN}}$.
A decreased N$_2$ concentration favors Ti-termination.
}
\end{figure}

\section{Results\label{sec5}}
Table~\ref{tab:molecules} lists the calculated atomization energies, 
bond lengths, 
and bond angles of molecules 
that are relevant for CVD of TiX.
For atomization energies, 
our calculated values are in very good agreement
for H$_2$ and N$_2$.
For other molecules, we find larger discrepancies, 
in particular for TiCl$_4$.
However, since the geometric properties
are in excellent agreement with experiment,
we suspect that these discrepancies
are mainly related to total energies of the isolated
atoms.
We therefore trust total energies of the listed molecules
as reference energies in the calculations of chemical potentials.

Figure \ref{fig:rCrit} displays the  
critical scaled deposition rates
$r^{\text{crit}}_{\text{TiX}}$ 
as functions of temperature
for two different pressures
and two different concentrations
of HCl in the supply gas.
We find that $r^{\text{crit}}_{\text{TiX}}$ first increases
with increasing temperature
and then approaches an asymptotic value.
At fixed temperature, $r^{\text{crit}}_{\text{TiX}}$ decreases
with increasing total pressure.
We also find a divergence in $\log_{10} r^{\text{crit}}_{\text{TiX}}(T)$
as the temperature decreases.
This divergence can be seen most pronounced for TiC
in the case where no HCl is supplied
(divergences of other graphs lie outside the plotted regime).
This divergence arises because lowering of 
the temperature below a certain value causes
$g_{\text{r}}^{\text{TiX}}$ to become positive
for all choices of  $r_{\text{TiX}}$.

\textit{TiC(111). }
Figure \ref{fig:HnucCVD}  reports the calculated 
free energies of reaction of Ti and C excess layers
as well as the free energy of reaction  
of one stoichiometric unit of bulk TiC.
We plot the free energies of reaction 
as  functions of the scaled 
reaction rate $r_{\text{TiC}}$
at fixed temperatures and pressures.
The scaled reaction rate is limited to right by its
critical value when dynamic equilibrium is reached,
see (\ref{eq:rCrit}).
We assume experimental values for the supply gas composition 
as stated in  Ref.~\onlinecite{Halvarsson1993177}.
In detail, we have
$c_{\text{TiCl$4$}}=0.04$, $c_{\text{CH$_4$}}=0.07$, $c_{\text{HCl}}=0.01$, and $c_{\text{H$_2$}}=1-\sum_i x_i$.

We find that $G_{\text{r}}^{\text{TiC/C}}>G_{\text{r}}^{\text{TiC/Ti}}$
in the entire range of the 
scaled deposition rate $r_{\text{TiC}}$.
Furthermore, the free energy of reaction associated
with a C-terminated TiC(111) surface is positive.
Thus, under the considered experimental circumstances,
the TiC(111) surface will be Ti terminated.

We have also tested the consequences of varying
deposition parameters.
With the chosen set of supply gases,
it was not possible to identify
a set of deposition parameters
that could lead to a favored C-termination
of the TiC(111) surface.

\textit{TiN(111).}
Figure \ref{fig:HnucCVDTiN}  reports the calculated 
free energies of reaction of Ti and N excess layers
as well as the free energy of reaction  
of one stoichiometric unit of bulk TiN.
In the upper panel,
we assume experimental values for the supply gas composition as stated in 
Ref.~\onlinecite{Larsson2002203}.
In detail, we have
$x_{\text{TiCl$4$}}=0.09$, $x_{\text{N$_2$}}=0.5$
and $x_{\text{H$_2$}}=1-\sum_i x_i$.

We find that the free energies of reactions
of both excess Ti and excess N can be negative
simultaneously.
Far from dynamic equilibrium, that is, 
for $r_{\text{TiN}}<<r_{\text{TiN}}^{\text{crit}}$,
we have $G_{\text{r}}^{\text{TiN/N}}>G_{\text{r}}^{\text{TiN/Ti}}$.
In this regime, a Ti-terminated surface is more favorable.
Close to dynamic equilibrium, that is, 
for $r_{\text{TiN}}\sim r_{\text{TiN}}^{\text{crit}}$,
we have $G_{\text{r}}^{\text{TiN/N}}<G_{\text{r}}^{\text{TiN/Ti}}$.
Furthermore, $G_{\text{r}}^{\text{TiN/Ti}}$ is positive
there.
In this regime, a N-terminated surface is more favorable.

In the lower panel we illustrate the effects of varying 
the deposition parameters.
Assuming an increased HCl content in the supply gas, 
balanced by the H$_2$ content,
we find that  $G_{\text{r}}^{\text{TiN/Ti}}$ turns positive
over the whole range of the scaled reaction rate.
Thus, we predict  a N-terminated TiN(111) surface.
Conversely, a decrease of the N$_2$ concentration
in the supply gas (balanced by H$_2$)
and a simultaneous increase of the deposition temperature,
to $T=1200^{\circ}$C
results into the favorization of  
a Ti-terminated surface.

\textit{In summary}, we find that,
under the conditions specified in  Ref.~\onlinecite{Halvarsson1993177},
the as-deposited TiC(111) surface will be Ti terminated.
Forcing a C-terminated surface by variation of process parameters
within the specified supply gas can not be achieved.
Under the conditions specified in Ref.~\onlinecite{Larsson2002203},
the TiN(111) surface can, in principle, be either Ti- or N-terminated,
depending on the ratio $r_{\text{TiN}}$ 
between the deposition and supply rate.
By introducing a slight amount of HCl into the supply gas,
a N-terminated surface will be forced.
With another choice of deposition parameters
a Ti-terminated surface will evolve.

\section{Discussion\label{sec6}}
The AIT-DG method that we have proposed in Sec.~\ref{sec:AIT-DG}
and illustrated in Sec.~\ref{sec5}
is designed to describe surface terminations
as a function of the growth environment.
Our predictions are based on calculations of 
the free energies of reaction $G_{\text{r}}$.
We motivate this criterion by the physical principle that all systems 
strive for the states of low energy,
given relevant boundary conditions.
The method applies in dynamic equilibrium
but also in scenarios where (dynamic) equilibrium is not maintained
but the system still settles into a steady state.
The method assumes that kinetic effects 
do not  lock the system into thermodynamically unstable morphologies.
It assumes in essence that a high deposition temperature 
causes a continuous annealing of  the emerging structure.

\subsection{Model limitations}
Kinetic-barrier effects certainly  exist and may prevent some systems
from actually reaching the structure and morphology
with the lowest thermodynamic energy.
\cite{PhysRevLett.83.2608}
Similar to the AIT-SE method,
our method neglects these barrier effects.
In principle, this approximation limits the applicability
of our method to high-temperature growth
where barriers become less important.
Nevertheless, 
kinetic effects can not be expected
to alter the predictions in cases similar to TiC.
This is because one of the free energies of reaction
is negative (Ti excess) while the other is positive (C excess).

Another obvious limitation is that of the ideal-gas approximation
from which we infer the temperature and pressure
dependence of the molecular chemical potentials.
Formally this approximation limits our description to 
low-pressure scenarios,
but the approach could easily be generalized by
a more refined description of the gas thermodynamics.

\subsection{Growth and rate equations}
Our method is capable of describing surface
terminations as they evolve during a growth process.
This is fundamentally different from 
predicting the termination of a surface 
that is in equilibrium with the environment.
As illustrated for TiN
in the top panel of figure \ref{fig:HnucCVDTiN},
the emerging surface termination depends on the 
the scaled reaction rate,
$r_{\text{TiX}}=R_{\text{TiX}}/R_{\text{sup}}$.
In such cases, our method should have clear advantages
over the AIT-SE method.
Use of our method requires only that
measurements of the scaled reaction rate
are made available.\cite{ratesComment}

We are dealing with an open system that has both
a gas flow and a consumption of gases during deposition.
In a closed system, 
where the number of gas-phase species is fixed,
the surface-gas system will approach 
equilibrium (unless there are important kinetic barriers).\cite{Demirel}
Similarily, the growing surface and the gas environment 
could be in a state of dynamic equilibrium.
In this case, the analysis in Sec.~\ref{sec:EquilibriumLimit} 
shows how the AIT-SE formalism could be extended to cover a 
description of the deposition.
This can even work for materials
that do not contain constituents A for which there exist
molecular counterparts A$_n$.\cite{O2vsN2}
Using the free surface energy as equilibrium predictor,
the chemical potentials of atoms in the gas can be
associated as an appropriate difference
between molecular chemical potentials.
An explicit example for TiC in an CH$_4$-TiCl$_4$-H$_2$
environment is given in Eq.~(\ref{eq:muAIG}).
However, since the chemical potentials of
atoms now depend on partial pressures of several
gas-phase species,
there is still a need to determine all relevant
partial pressures.
The rate-equation approach presented here
supplies the information that is necessary
for application of either the AIT-SE or the more general 
AIT-DG description of deposition growth.\cite{pressureDeviations}

We emphasize that growing systems,
in general, do not need to reach dynamic equilibrium,
but may be in any other steady state
(if steady-state is reached at all).
If dynamic equilibrium was always maintained, 
the scaled reaction rate $r_{\text{TiX}}$
would always assume its critical value,
that is, a maximum.
The absolute deposition rate could then be increased
to arbitrarily large values,
simply by increasing the supply rate at a fixed total pressure.
We believe that the assumption of dynamic equilibrium
is too optimistic for CVD systems, 
and we would not generally trust 
dynamic-equilibrium descriptions.
The here-proposed theory of deposition growth (AIT-DG)
applies also when the deposition 
and supply rates are independent
variables.

\subsection{Innovation potential}
Using the free energy of reaction 
for the formation of excess layers on a specified surface
we have shown for TiX(111) how surface terminations can be 
understood from knowledge of the details
of the CVD gas supply.
For TiN,  we have also illustrated that the method
can be used to guide the growth towards a desired surface composition.
That is, we have shown that we can modify the deposition environment
(supply gas composition,  deposition temperature, and  deposition pressure)
to allow either of the two  surface terminations
to emerge.

From this point of view our suggested method
has a potential for accelerating innovation.\cite{innovation}
In the case of TiX(111), the potential to design 
the surface termination may seem trivial.
The change from a Ti- to a N-terminated
TiN(111) surface that follows from raising the HCl concentration
in the supply gas could likely also be obtained by, 
terminating the deposition process
with an replacement of the TiCl$_4$-N$_2$-H$_2$ supply
gas with a pure N$_2$ supply gas.
We argue, however, that our method
is valuable for characterizing more complex 
materials with a larger variety of
possible surface terminations.
This is particularly true when we extend the application 
to understand binding  at interfaces 
that arise in deposition growth,\cite{rohrer_AIT_TiCAlumina}
for example, in the case of thin-film alumina
on TiC.\cite{rohrer_TiCAluminaStructure}

\section{Summary and conclusions\label{sec7}}
We present a method  based on \textit{ab-initio} calculations
supported by thermochemical data
to predict terminations of surfaces
as they emerge during high-temperature deposition growth
from a multicomponent gas,
for example chemical vapor deposition.
The method describes scenarios that are located
between a static equilibrium and
the fully non-equilibrium regime 
that requires a complete kinetic description. 
It relies on the basic thermodynamical principle
that all systems strive for low-energy states
subject to statistics and other boundary conditions.
Rate equations and associated steady-state solutions
play a central role in the calculation of 
free energies of reaction.
We use the latter to determine
the chemical composition of the outermost 
surface layer of the growing surface.

We illustrate the approach for TiX(111) (X = C or N).
For TiC, we predict a Ti-terminated surface
when fabricated under the conditions
stated in  Ref.~\onlinecite{Halvarsson1993177}.
For TiN, our predictions based on experimental conditions
reported in Ref.~\onlinecite{Larsson2002203}
are not so clear.
We find that the termination depends
on the ratio between the reaction rate and 
the rate at which the gas is supplied to the 
reaction chamber.\cite{ratesComment2}
In or close to dynamic equilibrium, 
we predict a N-terminated TiN surface.
Departing from dynamic equilibrium
will result in a Ti-terminated surface.
We also suggest deposition parameters
for which Ti- or N-terminated surfaces
can be achieved independent of
reaction and supply rates.

We also compare our method with the AIT-SE method.\cite{PhysRevB.62.4698,AIT_Scheffler} 
We show that the predictions of our method 
agree with those of the AIT-SE method
in the limit where (dynamic) equilibrium is maintained,
but also point out that this condition is not necessarily 
applicable to growth.
In a closed system (absence a gas flow),
there is an affinity to reach equilibrium.
In an open system (with the gas flow turned on), 
one could expect a corresponding affinity
to reach dynamic equilibrium.
However, as discussed in Sec.~\ref{sec6},
this expectation is likely too optimistic
for general CVD growth conditions.
In such cases, our method has clear advantages
since it does not \textit{a priori}
implement equilibrium constraints.

The choice of materials, in particular TiC,
also exemplifies the broader applicability of our method 
as compared to a strict implementation of the AIT-SE method.\cite{O2vsN2}
Our method is not formally limited 
to predictions of oxide surface stability 
(or as in  extensions of AIT-SE, 
to A$_m$B$_n$ compound surfaces 
for which there is an A$_k$ or B$_k$ counterpart in the gas phase).

Finally the capability of predicting different surface
terminations as a function of the deposition conditions
illustrate the  predictive power of our method.
It motivates continued development of 
\textit{ab-initio} thermodynamics,
also in a role to guide 
experimental optimization
of surface and interfacial structures.

\section*{Acknowledgments}
We thank T.S. Rahman and Z. Konkoli
for encouraging discussions during 
preparation of the manuscript.
Support from the Swedish National Graduate School in Materials Science
(NFSM),
from the Swedish Foundation for Strategic Research (SSF)
through ATOMICS,
from the Swedish Research Council (VR)
and from the Swedish National Infrastructure for Computing (SNIC)
is gratefully acknowledged.\\

\begin{appendix}

\section{Gibbs free energy of formation as predictor of surface terminations.}
Chemical reaction theory provides the basis to relate the probabilities
$P_{\text{A}}$ and $P_{\text{B}}$ for A and B surface termination 
to free energies of reaction $G_r$.
The surface transformations described by (\ref{eq:reactions})
can be cast into the following rate equations
for the probability of growing either an A- or B-terminated surface,
\begin{subequations}
\begin{align}
\partial_t P_{\text{A}}&=
-(\Gamma_b^{I}+\Gamma_f^{II}) P_{\text{A}} 
+(\Gamma_f^{I}+\Gamma_b^{II}) P_{\text{B}}\\
\partial_t P_{\text{B}}&=
(\Gamma_b^{I}+\Gamma_f^{II}) P_{\text{A}} 
-(\Gamma_f^{I}+\Gamma_b^{II}) P_{\text{B}}.
\end{align}
\end{subequations}
The steady-state solution is
$P_{\text{A}}/P_{\text{B}}=
(\Gamma_f^{I}+\Gamma_b^{II})/
(\Gamma_b^{I}+\Gamma_f^{II})$
where
$\Gamma_{\{f,b\}}^{\{I,II\}}$ are
forward (f) and backward (b) reaction rates
identified in reaction (\ref{eq:R1}) and (\ref{eq:R2}).
These rates can be expressed as the products of 
the corresponding microscopic rate constants
$k_{\{f,b\}}^{\{I,II\}}$,
and of an environment-specific product 
$Q_{\{f,b\}}=\Pi_i[\text{X}_i]^{\nu_{\{f,b\},i}^{\{I,II\}}}$, Ref.~\onlinecite{Demirel} and \onlinecite{Tschoegl}.
Here [X$_i$] denotes the concentration of molecule X$_i$
and $\nu_{\{f,b\},i}^{\{I,II\}}$ is the (positively counted) stoichiometric coefficient 
as reactant (f) and product (b) in reaction  (\ref{eq:R1}) or (\ref{eq:R2}).
We note that $Q_{\{f,b\}}$ is strictly speaking a product of activities
from all parts in the reactions (\ref{eq:R1}) and (\ref{eq:R2}),
but for surfaces (or solid) the activity can be approximated by unity.

The general relation $\beta G_r=-\ln K + \ln Q$,
with $\beta$ the inverse temperature,
relates $G_r$ to the equilibrium constant $K=k_f/k_b$ and $Q=Q_b/Q_f$.
We have  
$Q_{\{f,b\}}^{\{I,II\}}=\Gamma_{\{f,b\}}^{\{I,II\}}/k_{\{f,b\}}$
and find 
$\exp\left(-\beta\left[G_r^{\text{A}}-G_r^{\text{B}}\right]\right)
=\Gamma_f^{I}\Gamma_b^{II}/(\Gamma_b^{I}\Gamma_f^{II})$.

Dynamic equilibrium is characterized by 
$G_r^{\text{A}}+G_r^{\text{B}}=0$
or equivalently $\Gamma_f^{I}\Gamma_f^{II}/(\Gamma_b^{I}\Gamma_b^{II})=1$,
from which it follows that 
\begin{align}
P_{\text{A}}/P_{\text{B}}\big|_{\text{dyn.eq.}}=\exp\left(-\beta\left[G_r^{A}-G_r^{B}\right]/2\right).
\label{eq:probabilities}
\end{align}
Away from dynamic equilibrium, this relation may not hold exactly.
However, whithin a broad range of nonequilibrium conditions
we may still approximate the arithmetric mean [$\langle x,y\rangle_{\text{AM}}=(x+y)/2$]
in  $P_{\text{A}}/P_{\text{B}}=\langle\Gamma_f^{I},\Gamma_b^{II}\rangle_{\text{AM}}/
\langle\Gamma_b^{I},\Gamma_f^{II}\rangle_{\text{AM}}$
by the geometric mean [$\langle x,y\rangle_{\text{GM}}=(xy)^{1/2}$] expressing  
$\exp\left(-\beta\left[G_r^{A}-G_r^{B}\right]/2\right)=\langle\Gamma_f^{I},\Gamma_b^{II}\rangle_{\text{GM}}/\langle\Gamma_b^{I},\Gamma_f^{II}\rangle_{\text{GM}}$.
This suggests that the evaluation (\ref{evaluation})
remains a good approximate measure of relative surface termination also
out of equilibrium.

\end{appendix}


\begin{references}
\bibitem{stoichiometry}
We note, however, that both are stoichiometric
films in the limit of a large thickness.
\bibitem{bulkFreeEnergy}
The (total) free energy per bulk unit is conveniently defined
as 
$g_{\text{TiX}}=\lim_{n\rightarrow\infty}
G_{\text{TiX}}(n+1)-G_{\text{TiX}}(n)
$
where $n$ is the number of bilayers in the
TiX slab.
Since the bulk energy does not depend on the actual surface
termination, we could also 
replace $G_{\text{TiX}}(n+1)$
by $G_{\text{TiX/Ti}}(n+1)$ or 
$G_{\text{TiX/X}}(n+1)$
if we simultaneously replace
$G_{\text{TiX}}(n)$
by $G_{\text{TiX/Ti}}(n)$ or 
$G_{\text{TiX/X}}(n)$ respectively.
\bibitem{EquivalentDefinition}
We note that we could have chosen
non-stoichiometric slabs as initial 
configurations of the surface and adsorb
additional layers on one of the two
equivalent sides of this slab.
In this case, we need to use a modified
version of the definition
of  $G_{\text{r}}^{\text{TiX/Ti}}$.
We emphasize, however, that the results presented
here are independent of the choice
of the definition of $G_{\text{r}}^{\text{TiX/Ti}}$,
that is, both definitions are equivalent.
\bibitem{SurfaceVibComment}
We are not aware of determinations of mean frequencies of Ti and X in TiX (X = C,N)
which could directly enable an estimate of the vibrational contributions 
similar to that given in Ref.~\onlinecite{AIT_Scheffler}.
Because there is a large difference in masses between Ti and X
one can expect a larger correction in the vibrational free energy
for the lighter X.
However, differences in such vibrational contributions will only shift 
the crossing points in the free energies of reaction
and therefor not alter qualitative predictions
or the value of the presented method.
\bibitem{O2vsN2}
The AIT-SE formalism is essentially designed
for oxides in equilibrium with an O$_2$-dominated environment.
Here, we consider TiX in a much more complicated environment.
However, the above definitions, steps, and conclusions readily
carry over to growth of an oxide
in an environment that is more complex
than a pure or O$_2$-dominated environment.
\bibitem{ratesComment} 
In practice, measuring
the exhaust rate $R_{\text{exh}}$ 
(as a function of the supply rate $R_{\text{sup}}$)
that is required to keep the total deposition pressure
constant suffices to determine the scaled deposition rate.
\bibitem{pressureDeviations}
We note that rate equations may be required
even in cases where the relevant atomic chemical potential
can be determined from that of gases that are supplied
and for which the concentrations in the supply gas is known.
Equation~(\ref{eq:SSpressures}) 
clearly shows that the steady-state concentrations
will in general differ from the concentrations in the supply gas.
Using experimental deposition parameters for TiN,
we find that deviations in the steady-state pressure
$p_i$ from those that are expected from
the concentration in the supply gas
$p_i^0=c_i\cdot p$ can become as large as 20\%.
\bibitem{ratesComment2}
This ratio is, in principle,
accessible by monitoring
the supply and exhaust rate
at a fixed pressure.
\end{references}
\end{document}